\documentclass{PoS}

\def\eiso{E$_{\rm iso}$}
\def\epeak{E$_{\rm peak}$}

\title{Gamma-Ray Bursts, witnessing the birth of stellar mass black holes}

\ShortTitle{Gamma-Ray Bursts, witnessing the birth of stellar mass black holes}

\author{\speaker{Jean-Luc Atteia}\\ %         \thanks{}\\
        IRAP; Universit\'e de Toulouse; UPS-OMP; CNRS; 14, avenue Edouard Belin, F-31400 Toulouse, France\\
        E-mail: \email{jean-luc.atteia@irap.omp.eu}}

\abstract{Gamma-ray bursts are associated with catastrophic cosmic events. They appear when a new black
hole, created after the explosion of a massive star or the merger of two compact stars, quickly
accretes the matter around it and ejects a transient relativistic jet in our direction.
This review discusses the various types of gamma-ray bursts, their progenitors, their beaming and their rate in the local universe.
We emphasize the broad astrophysical interest of GRB studies, and the crucial role of high-energy satellites 
as exclusive suppliers of GRB alerts and initial locations.}

\FullConference{ An INTEGRAL view of the high-energy sky (the first 10 years) - 9th INTEGRAL Workshop and celebration of the 10th anniversary of the launch\\
		 15-19 October 2012\\
		 Biblioth\`eque Nationale de France, Paris, France}

\begin{document}

\section{Introduction}
\label{intro}
\textit{INTEGRAL} has spent more than ten years in space, observing the gamma-ray sky, scanning the galaxy in the light of gamma-ray lines, tracing cosmic rays on their travel from their sources, looking for traces of dark matter, and exploring regions of the universe where large amounts of energy are released violently. Gamma-ray bursts (GRBs) belong to this last category, being among the most transient, the most luminous, and the most distant sources detected by \textit{INTEGRAL}. 

\textit{INTEGRAL} has the capability of detecting GRBs within the fields of view of SPI and IBIS, its two wide-field instruments, but also on the 5200 $\rm{cm}^2$ effective area of the BGO anticoincidence shield of SPI. The data collected by \textit{INTEGRAL} instruments are continuously transmitted to the ground, allowing fast GRB detection and localization. The value of such fast localizations was realized early, leading Mereghetti et al. to develop IBAS, the \textit{INTEGRAL Burst Alert System} \cite{Mereghetti2003}, which holds the record of the fastest GRB alerts with delays as short as 8-10 seconds after the trigger. Since then, \textit{INTEGRAL} has permitted numerous interesting discoveries on GRBs which are summarized in this volume by D. G\"otz \cite{G\"otz2012}. In this paper we discuss the origin of GRBs, the nature of their progenitors, and their connection with the birth of stellar mass black holes.

\section{What are gamma-ray bursts?}
\label{grb}

\begin{figure}[htbp]
\begin{center}
\includegraphics[width=15cm]{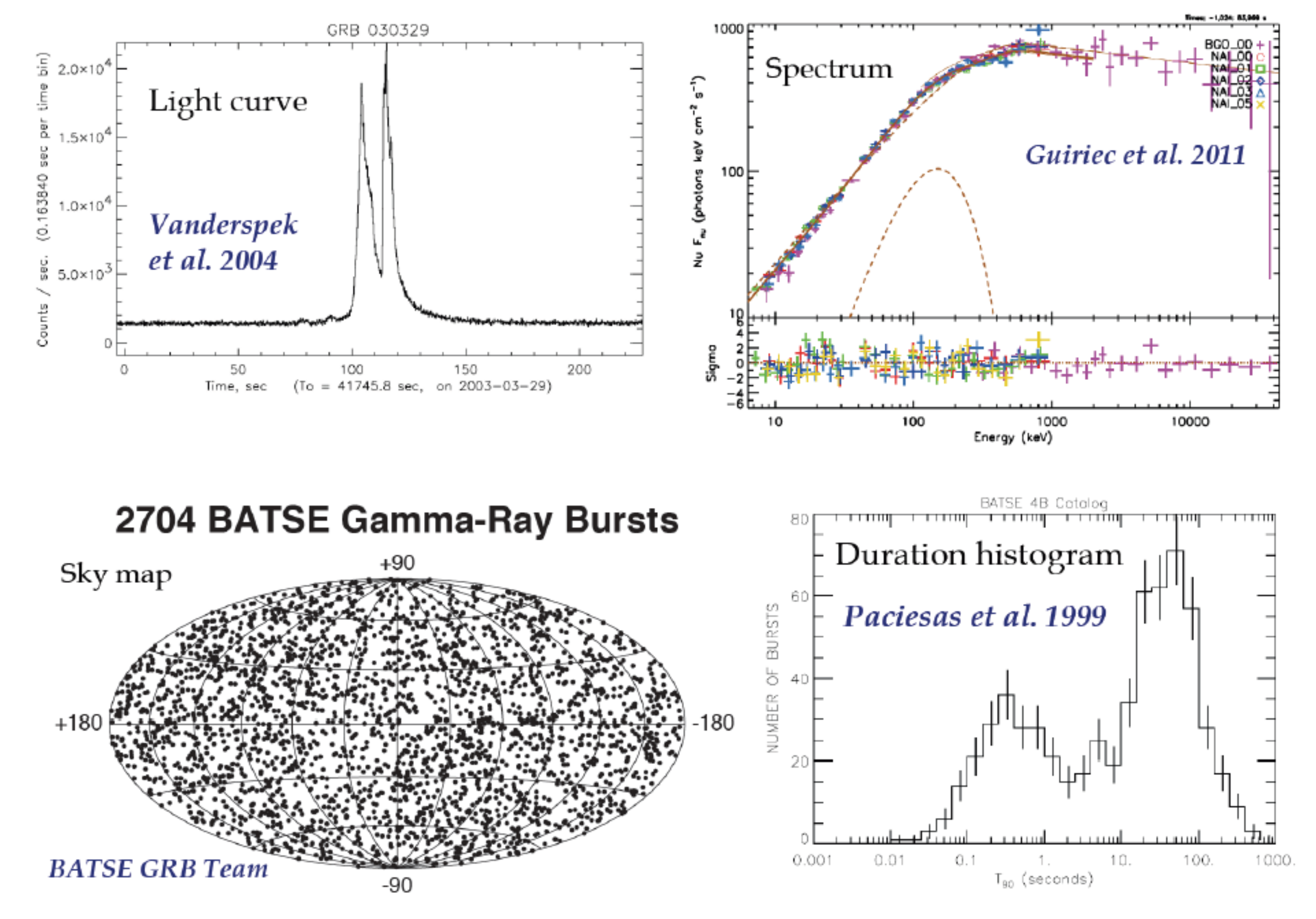}
\caption{Some properties of prompt GRB emission. \textit{Upper left:} Light-curve of the bright GRB 030329 in the range 7-400 keV \cite{Vanderspek2004}. \textit{Upper right:} $\nu {\rm F} \nu$ spectrum of GRB 100724B \cite{Guiriec2011}. \textit{Lower left:} Sky distribution showing the isotropy of GRBs \cite{Batse}. \textit{Lower right:} Duration histogram of prompt high-energy emission, showing the long and short GRB populations \cite{Paciesas1999}}
\label{fig1}
\end{center}
\end{figure}

Gamma-ray bursts (GRBs) are bright flashes of high-energy photons coming from very distant galaxies \cite{Vedrenne2009}. Figure \ref{fig1} shows the main properties of prompt GRB emission: light-curves exhibiting a flash of hard X-ray photons lasting from tens of milliseconds to hundreds of seconds, $\nu {\rm F} \nu$ spectra peaking around \epeak\ $\sim 100$ keV, and sources isotropically distributed on the sky \cite{Batse}. The duration histogram shows two peaks (with maxima around 0.3 and 40 seconds) suggesting the existence of two classes, respectively called short and long GRBs. 
High-energy satellites play an essential role in the detection and classification of GRBs. The Large Area Detectors of \textit{BATSE}, which have worked more than 9 years have shown that about 1000 GRBs cross the solar system each year (hereafter we call these bursts "classical GRBs", in contrast with X-Ray Flashes or low-luminosity GRBs described below). Since then, instruments like \textit{BeppoSAX} WFC and \textit{HETE-2} WXM have extended GRB detection to soft X-rays, revealing the existence of very soft GRBs (\epeak\ $\le 30$ keV), which have been called X-Ray Flashes (XRFs) \cite{Heise2003, Kippen2003, Barraud2003, Sakamoto2005}. XRFs are closely connected with GRBs, except for their \epeak\ and their fluence, which are 10-50 times smaller than regular GRBs. 
Rarely a low-luminosity GRB is detected in the local universe (z $< 0.1$). Assuming isotropic emission, it radiates an energy which is $10^3 - 10^4$ times fainter than classical GRBs. These GRBs are called low-luminosity GRBs, they seem to dominate the GRB population, but their connection with the population of bright classical GRBs is poorly understood.

After nearly three decades where only the prompt emission was detected, \textit{BeppoSAX} has permitted the discovery of the afterglows thanks to the distribution of arcminute positions within few hours of the burst \cite{Costa1997,vanParadijs1997}. Then, \textit{HETE-2}, \textit{INTEGRAL} and \textit{Swift} have set a new standard, distributing arcminute locations within seconds of the GRB. With \textit{Swift}, these positions are improved to several arcseconds, thanks to the localization of the X-ray afterglow with the XRT. The remarkable combination of sensitivity, speed and accuracy achieved by \textit{Swift} for GRB localization has permitted measuring more than 200 GRB redshifts and host galaxies.

GRBs are followed by quickly decaying afterglows, which are visible at all wavelengths. These afterglows permit to locate GRBs accurately, to find their host galaxies, to measure their distance and energetics, and to study their interaction with their environment. The location of \textit{long GRBs} in star forming regions and the clear association of several nearby GRBs with hypernovae \cite{Galama1998,Hjorth2003,Stanek2003} have provided strong evidence that long GRBs are connected with the end of life of massive stars. The supernovae associated with GRBs have special properties, they are hypernovae (supernovae with large kinetic energy of the ejecta) of type Ibc (without hydrogen and helium lines, indicating that they had ejected their envelope before the explosion). The favored models, invoke a newly born magnetar or the fast rotating core of a massive star collapsing into a black hole. The origin of \textit{short GRBs} is less constrained but their frequency, their occurrence in all types of galaxies and their energetics suggest an association with mergers of compact stars (two neutron stars or a black hole and a neutron star). The coalescence of the two stars results in a system composed of a black hole quickly accreting a torus of residual material.

\section{GRBs are rare and powerful}
% ou... \section{GRB statistics}
\label{jets}

\begin{figure}[ht]
\centering
\begin{minipage}{.49\textwidth}
%\hspace{-0.5cm}
\vspace{1.8cm}
\includegraphics[width=\textwidth]{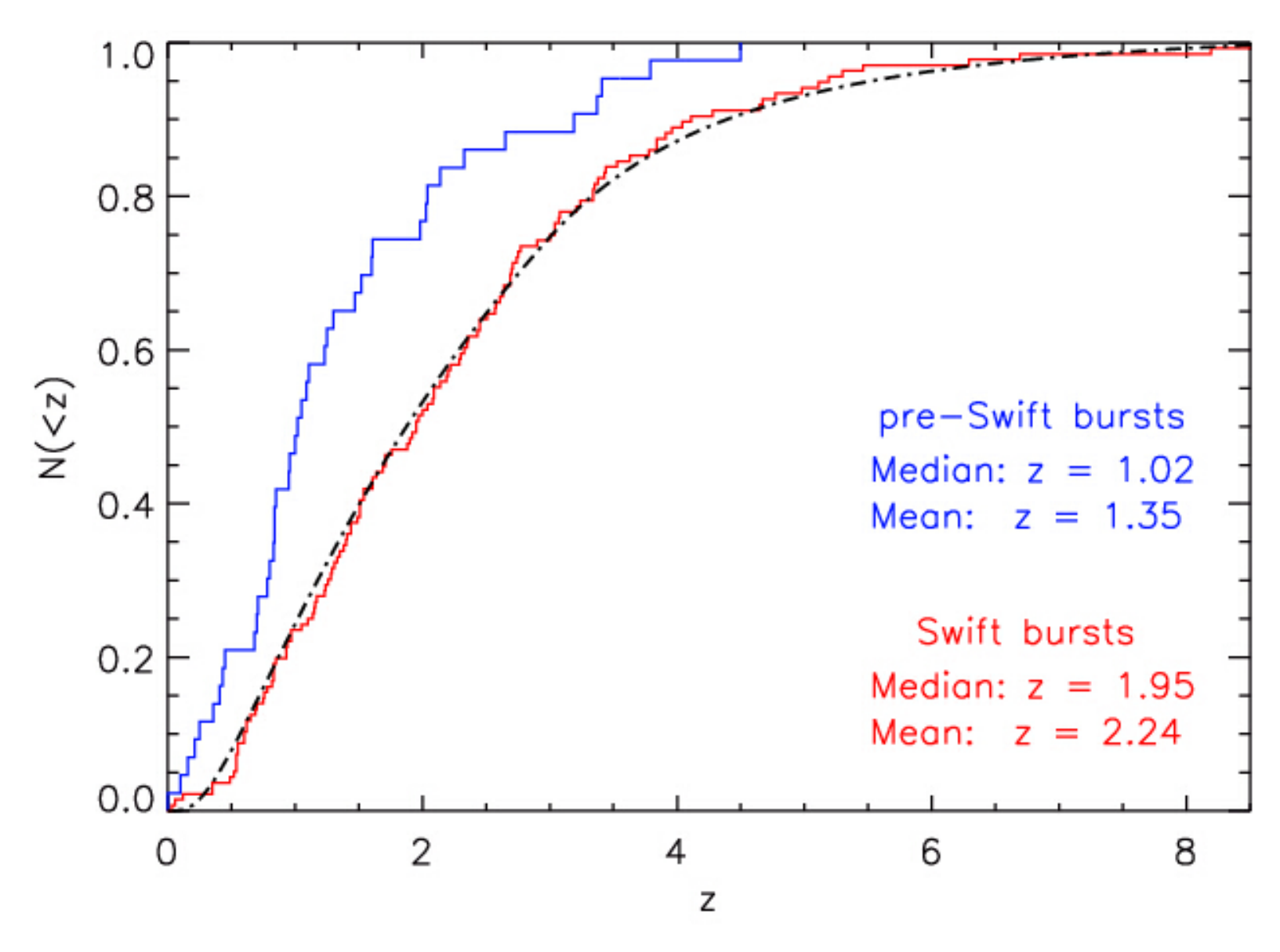}
\end{minipage}
\begin{minipage}{.49\textwidth}
\includegraphics[width=\textwidth]{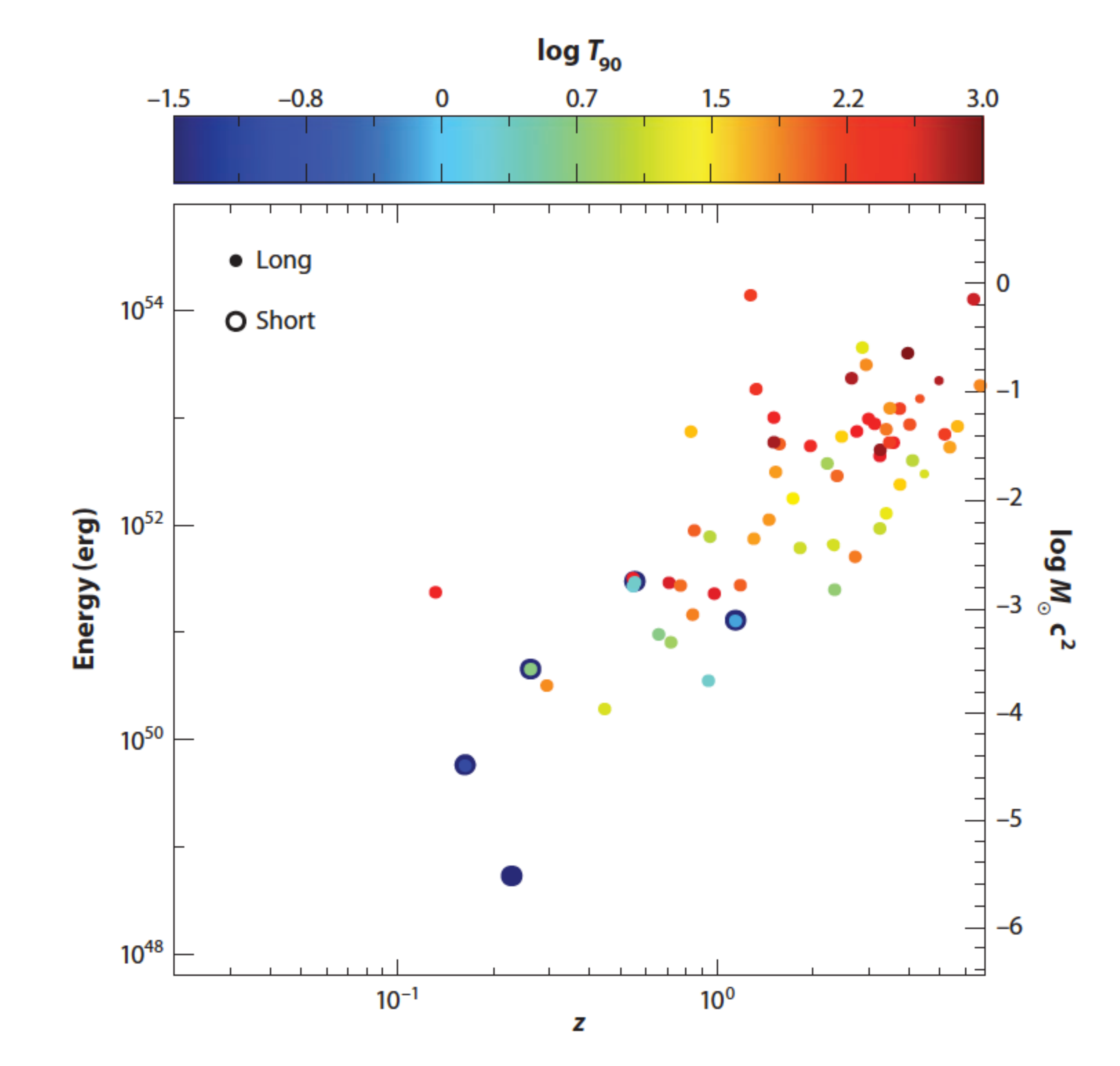}
\end{minipage}
\caption{\textit{Left:} Distribution of redshifts of \textit{Swift} and pre-\textit{Swift} GRBs \cite{Jakobsson2012}. \textit{Right:} Distribution of \textit{Swift} GRBs in a redshift-energy plane \cite{Gehrels2009}, showing the existence of bursts releasing up to $10^{54}$ erg in $\gamma$-rays.}
\label{fig2}
\end{figure}

\textit{Long GRBs. }
The measure of GRB redshifts (see figure \ref{fig2}) has evidenced the extreme luminosity of these sources: long GRBs have been detected out to very high redshifts (z $> 8$), and if one assumes isotropic radiation, some of them would release up to $10^{54}$ \textit{in $\gamma$-rays}. This is much more than the kinetic energy of a supernova and it is generally admitted that long GRBs are jetted with a beaming factor in the range 100 - 1000 ($\theta_{jet}$ = 2.5$^\circ$ to 8$^\circ$) and an energy budget in $\gamma$-rays in the range $10^{51}$ - $10^{52}$ erg. The standard fireball model predicts an achromatic steepening of the light-curve when the Lorentz factor falls below the opening angle of the jet ($\Gamma < 1/\theta_{jet}$). Such breaks have been observed in some GRBs after 1 or few days, leading to opening angles of several degrees, but it is not known how much these measures are affected by selection effects (late-time breaks are difficult to measure because the flux of the afterglow is low).
Radio observations performed several months after the burst, when the jet is sub-relativistic and its energy is randomized, have shown that the kinetic energy of the jet clusters around few $10^{51}$ erg \cite{Berger2004}. Since the gamma-ray luminosity is not expected to exceed by orders of magnitude the kinetic energy of the jet measured at late time, these measurements require a strong beaming of the brightest GRBs. The opening angle of GRB jets remains poorly constrained, but there is hope for progress in the coming years with the advent of deep optical and radio surveys, which will have the sensitivity to detect \textit{orphan afterglows}. Orphan afterglows are due to GRBs which are not pointing towards the Earth, they become visible after few days in the optical when the bulk Lorentz factor of the jet becomes smaller than the opening angle of the jet, allowing lateral expansion of the jet. Radio afterglows remain visible during several weeks or months, until the jet becomes subrelativistic ($\Gamma < 1$) \cite{Soderberg2010}.

The beaming factor impacts the estimates of GRB spatial density. The observed local rate of classical GRBs is $\sim$1 Gpc$^{-3}$ yr$^{-1}$, after correction for beaming this rate increases to $100-600$ Gpc$^{-3}$ yr$^{-1}$ \cite{Guetta2007}. In comparison the rate of type II supernovae is $\sim$10$^5$ Gpc$^{-3}$ yr$^{-1}$, about ten times the rate of hypernovae ($\sim$10$^4$ Gpc$^{-3}$ yr$^{-1}$). Taking these numbers at face, only few percent of hypernovae produce a GRB. We discuss in the next section the parameters which could lead a massive star to emit a GRB at the end of its life.

\paragraph{}
\textit{Short GRBs. }
The observed rate of short GRBs is $\sim$10 Gpc$^{-3}$ yr$^{-1}$ \cite{Nakar2006,Coward2012}. The beaming factor of short GRBs is less constrained than for long GRBs but it is usually admitted that short GRBs are less beamed, with beaming factors in the range 10 - 100 ($\theta_{jet}$ = 8$^\circ$ to 26$^\circ$) \cite{Nakar2006,Coward2012}. Interestingly, this leads to a rate corrected for beaming which is comparable with the rate of long GRBs, $100-1000$ Gpc$^{-3}$ yr$^{-1}$. This is also comparable to the expected frequency of mergers of double neutron stars computed from available binary pulsar observations and evolutionary calculations of double neutron star formation \cite{Belczynski2008}. In the present context, the fraction of binary neutron star mergers which gives rise to short GRBs could thus range from several \% to nearly 100\%.

\paragraph{}
\textit{Low-luminosity GRBs. }
A handful of low-luminosity GRBs (LL GRBs) have been detected in the local universe (z $<$ 0.1) \cite{Galama1998, Campana2006, Pian2006, Soderberg2006}. These bursts have been associated with hypernovae of type Ibc, confirming their close connection with their luminous cousins. Given their proximity, LL GRBs permit detailed studies of their progenitors and regions of production. Extrapolating from the few events detected, the rate of LL GRBs is 100-500*$f_b$, where $f_b$ is the beaming factor of LL GRBs, ranging from 1 to 10. Assuming that all LL GRBs are associated with hypernovae implies that the rate of LL GRBs must not exceed the rate of hypernovae ($\sim$10$^4$ Gpc$^{-3}$ yr$^{-1}$). This has two consequences: first, low-luminosity GRBs must have a small beaming factor ($\lesssim$10), second, the \eiso\ distribution cannot extend much below $10^{48}$ erg, the \eiso\ of GRB 980425. 
While the link between LL GRBs and classical GRBs is not elucidated, it is generally assumed that they constitute different populations \cite{Soderberg2006,Daigne2007}. For instance, some authors have suggested that LL GRBs could be associated with magnetars \cite{Mazzali2006} instead of accreting black holes.

\section{GRB progenitors}
\label{progenitors}

%\paragraph{}
\textit{Long GRBs and LL GRBs. }
Long GRBs and LL GRBs are due to massive star explosions producing hypernovae with transient relativistic jets. The enormous energy released by these events appears in various forms: the kinetic energy of the hypernova, a large amount of Nickel 56, and a relativistic jet powered by a central engine active during several seconds. The energetics may be dominated by the jet (GRBs) or by the kinetic energy of the hypernova (LL GRBs). The presence of a jet in nearby hypernovae can be inferred with radio observations which trace the relativistic outflow, even when it is not directed toward us. The vast majority of Type Ibc hypernovae show no evidence of a relativistic jet, and it seems that at most a few percent of SN Ibc produce GRBs \cite{Berger2003, Soderberg2010}. This raises the question of the nature of the central engine and the conditions needed for its existence. 

The association of GRBs with massive stars and the well-known ability of black holes to accelerate relativistic jets (in active galactic nuclei and galactic microquasars) have led to models explaining the production of GRBs by cores of massive stars which collapse into black holes. Numerical simulations have shown that the stellar core must have a large angular momentum to permit the building up of an accreting torus, and that, under some conditions, the jet accelerated by the black hole can pierce the stellar envelope and escape into space (e.g. \cite{Woosley1999, Zhang2004}). 
This model raises numerous interesting questions, which concern the physical processes at work close to the black hole and in the jet, the possibility to keep a large angular momentum in the stellar core while ejecting the envelope of the progenitor, the possibility of using the prompt emission to infer the nature and the properties of the central engine, the afterglow evolution which is most often successfully fit by a jet interacting with a constant density medium and more rarely with the stellar wind of a massive star... 
These questions show that we are far from understanding these complex sources. It is essential to gather more information on GRB progenitors, to understand better the role of binarity and metallicity in the final fate of massive stars, and to clarify if short-lived magnetars or quark stars are formed during the collapse. Nearby GRBs play a crucial role in this quest.

\paragraph{}
\textit{Short GRBs. }
Unlike long GRBs, short GRBs are not associated with young massive stellar populations. Their occurrence in all types of galaxies, their lower energy output, and their rate are compatible with their origin in mergers of compact stars (binary neutron stars or a neutron star with a black hole). Their distance and frequency make them privileged sources of gravitational waves (GW) for advanced gravitational wave detectors. It is not unrealistic to think that the first gravitational wave signal of astrophysical origin will result from the joint detection of a short GRB in gamma-rays and GW. In this context it is very important to characterize short GRBs better, for instance by understanding their beaming pattern (which will tell us the fraction of binary mergers that produce GRBs) or looking for possible correlations between the properties of short GRB and the properties of their host galaxies.

\section{Conclusion}
\label{conclusion}

In this short review, it was only possible to discuss one facet of GRBs, their origin and the nature of their progenitors. In general GRB studies encompass a much broader range of subjects, highly relevant to today's astrophysics: the physics of the prompt emission, GRBs as probes of the distant universe, GRBs as tracers of the history of star formation, GRBs as potential tools to detect the first generation of stars if they explode, GRBs as cosmic rulers for cosmography... These questions show the importance of continuing observing GRBs. Fortunately, \textit{INTEGRAL} and \textit{Swift} are operating flawlessly and should be able to provide GRB alerts for the years to come. 

In the mid-term future, it is expected that GRBs will continue to play a major role in the progress of astrophysics, thanks to new instrumentation becoming available on the ground (Advanced LIGO and VIRGO, E-ELT, LSST, SKA, CTA, HAWC) and in space (JWST), which will allow multi-messenger studies of GRBs, and the detection of their afterglows and hosts out to the early times of star and galaxy formation (when the reionisation of the universe took place). This will be the time of SVOM, which is to be launched in the 2nd half of the decade. SVOM will provide complete data on GRBs during the burst and its afterglow \cite{Godet2012}. It will explore the realm of soft GRBs above 4 keV, and the prompt optical emission with a good sensitivity. Thanks to a pointing strategy optimized for ground follow-up, to the good sensitivity of its visible telescope, and to NIR follow-up on the ground, more than half of SVOM GRBs will have a redshift. 
This timeframe will also see the opening of new windows on GRBs with the advent of deep synoptic surveys allowing the discovery of optical or radio afterglows from the ground without the need for space alerts from high-energy satellites. Even if these windows will not permit measuring the energy of GRBs (because we do not see the prompt emission), they will undoubtedly give us new lights on these violent phenomena.

\end{document}